\documentclass[aps,prl,twocolumn,groupedaddress,showpacs]{revtex4}
\usepackage{amssymb}
\usepackage{amsmath}
\usepackage{cases}
\usepackage{graphicx,color}
\definecolor{r}{rgb}{1,0,0}   
\definecolor{g}{rgb}{0,1,0}   
\definecolor{b}{rgb}{0,0,1}
\begin{document}

\title{Effect of interstitial fluid on the fraction of flow microstates\\ that precede clogging in granular hoppers}

\author{Juha Koivisto and Douglas J. Durian}
\affiliation{Department of Physics and Astronomy, University of Pennsylvania, Philadelphia, Pennsylvania 19104-6396, USA }

\date{\today}

\begin{abstract}
We report on the nature of flow events for the gravity-driven discharge of glass beads through a hole that is small enough that the hopper is susceptible to clogging.  In particular, we measure the average and standard deviation of the distribution of discharged masses as a function of both hole and grain sizes.  We do so in air, which is usual, but also with the system entirely submerged under water.  This damps the grain dynamics and could be expected to dramatically affect the distribution of the flow events, which are described in prior work as avalanche-like.  Though the flow is slower and the events last longer, we find that the average discharge mass is only slightly reduced for submerged grains.  Furthermore, we find that the shape of the distribution remains exponential, implying that clogging is still a Poisson process even for immersed grains.  Per Thomas and Durian, Phys. Rev. Lett. {\bf 114}, 178001 (2015), this allows interpretation of the average discharge mass in terms of the fraction of flow microstates that precede, i.e.\ that effectively cause, a stable clog to form.  Since this fraction is barely altered by water, we conclude that the crucial microscopic variables are the grain positions; grain momenta play only a secondary role in destabilizing weak incipient arches.  These insights should aid on-going efforts to understand the susceptibility of granular hoppers to clogging.
\end{abstract}


\maketitle


\section{Introduction}

The flow of grains in an hourglass or hopper is an iconic granular phenomenon, strikingly different from the gravity-driven flow of fluid from a small hole in the bottom of a bucket \cite{NeddermanSavage}.  At the top free surface, the grains are not level but form a conical depression.  Below the hole, the stream of grains fans out and does not break up into droplets because there is no surface tension.  And the growing mass of grains collected underneath is not level, but forms a conical pile down which the added grains avalanche.  Another striking difference is that the discharge rate of grains is constant, as long-described by the empirical Beverloo equation \cite{NeddermanSavage, Beverloo}, and sometimes can even increase with time \cite{Wilson2014, SandsOfTime}; by contrast, the fluid discharge rate always decreases with time as the bucket empties and the gravitational pressure head goes down.  But an even bigger difference is that grains can clog \cite{HerrmannEPJE00, ToPRL01, MazaPRE03}.  Though the flow may appear smooth, it can suddenly and unexpectedly halt due to the formation of a mechanically stable arch or dome of grains spanning the hole.  How to predict the susceptibility of a given granular system to clogging \cite{ZuriguelSM2016}, and how to anticipate that a clog is about to form \cite{TewariSM2013}, are active research topics.

Clogging is a generic feature of granular hopper flow, as reviewed in Refs.~\cite{ZuriguelPIP14, Zuriguel2014}, which happens less frequently for larger holes and is unavoidable for holes smaller than about 4-5 grains across.  Details depend on grain shape, e.g.\ \cite{Saraf, Thomas2013, BehringerEPS16, StannariusNJP16}, and similar phenomena arise in other contexts ranging from transport in electronic \cite{OlsonReichhardt2013} and particulate \cite{CoussotPRL07, Reichhardts2016} systems with spatially-distributed pinning sites to grains in channels and pipes \cite{WyssPRE06, Janda2015}, grains driven by fluid flow \cite{GuariguataPRE2012, LafondPRE2013}, and even grains with brains: pedestrians \cite{Garcimartin2016}, traffic \cite{Nagatani2002}, and livestock \cite{Sheep2015}.  For non-cohesive compact grains, in air or vacuum, there is general agreement that clogging statistics are Poissonian \cite{ToPRL01, MazaPRE03, MazaPRE05, Tang09, Thomas2013}.  Namely, there is an exponential distribution of flow times, and hence also an exponential distribution for the amount of material discharged between successive clogs.  Thus there is a well-defined average ``avalanche" size, as measured either from the average flow duration $\langle \tau\rangle$ or from the average mass $\langle m \rangle$ discharged before a clog occurs.  These are related by $\langle m \rangle = \rho A v \langle \tau\rangle$ where $\rho$ is the mass density of the packing, $v$ is the exit speed of grains at the hole, $A=\pi(D/2)^2$ is the hole area, and $D$ is the hole diameter.

\begin{figure*}[ht]
\includegraphics[width=5in]{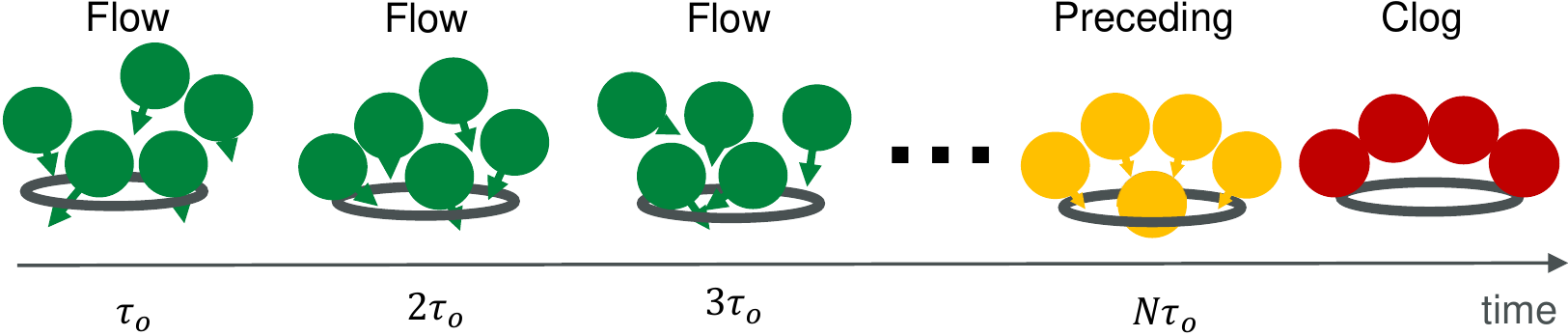}
\caption{When time advances by one sampling time, $\tau_o$, there is a new configuration of grains near the hole and a new chance to clog.  In this cartoon, $N$ distinct flow states are sampled prior to the formation of a stable clog.  The fraction of microstates that similarly cause clogging is $F=1/\langle N\rangle$, where $\langle N\rangle$ equals the average flow duration divided by $\tau_o$.}
\label{cartoon}
\end{figure*}

A crucial open question is whether or not a sharp clogging transition exists:  Is there a critical hole diameter $D_c$ above which the system will never clog, or instead does clogging become so improbable for larger holes as to be essentially unobservable on human time scales?  This is difficult to answer definitively by experiment or simulation because $\langle m\rangle$ grows very rapidly with hole size, e.g. by five order of magnitude as the hole diameter increases by a factor of three.  In particular, data can be equally-well described by both exponential \cite{ToPRL01, Janda08, Thomas2015} and diverging power-law \cite{MazaPRE03, Janda08, Thomas2013, Thomas2015} forms:
\begin{eqnarray}
  \langle m \rangle &=& m_g \exp[c\,(D/d)^3 + b] \label{mexp} \\
  \langle m \rangle &=& m_g / [\alpha (D_c - D)/d]^\gamma \label{mdiv}
\end{eqnarray}
where $m_g$ is the average grain mass, $d$ is the average grain diameter, and $\{c, b, \alpha, \gamma\}$ as well as the putative critical hole size $D_c$ are fitting parameters. This is also true for two-dimensional hoppers, where the average mass grows as either a critical power-law or as an exponential in $(D/d)^2$ \cite{ToPRL01, Janda08}.  While the competing fits may be equally good, we prefer exponential form for several reasons.  As negative points against the power-law, fitting results for the exponent $\gamma$ are oddly large and published results vary widely.  There is no established theory for the expected value of $\gamma$, or for the putative critical hole size $D_c$.  In principle, stable arches can be constructed of arbitrarily large size.  Furthermore, in spite of explicit searches, no critical signature such as a kink in either the average discharge rate \cite{Thomas2013} or in grain velocity fluctuations \cite{Thomas2016} has been found that could be used to locate $D_c$ as the hole size is decreased toward the putative transition from above.  Lastly, the exponential form and its dependence on dimensionality both follow naturally from a simple model based on consideration of clogging as a Poisson process in which microstates are randomly sampled \cite{Thomas2015}.

The first step in the model of Ref.~\cite{Thomas2015}, and the inspiration for the present paper, is the realization that the fraction $F$ of all accessible microstates that precede a clog can be found from measurement of $\langle m\rangle$.   Since clogging is a Poisson process, the act of flow can be interpreted as bringing new configurations into the region of the hole, so that with time different configurations are sampled at random until one arises that causes a clog to form.  We call $\ell$ the sampling length, which is how far the grains near the hole must flow in order to produce a new configuration.  It is of order one grain diameter, and the corresponding sampling time is $\tau_o = v \ell$.  This is illustrated in Fig.~\ref{cartoon}.  The average discharge mass may then be rewritten as $\langle m \rangle = \rho A v \langle \tau\rangle = \rho A \ell \langle \tau \rangle /\tau_o$.  In this expression, we recognize $\langle \tau\rangle / \tau_o$ as the number of distinct configurations sampled in the average flow event, and hence $F=\tau_o/\langle \tau \rangle$ as the fraction of flow configurations that precede a clog.  Thus the fraction of flow microstates that cause a clog is
\begin{equation}
	F = \rho A\ell/\langle m\rangle,
\label{FfromM}
\end{equation}
and can be deduced from measurement of $\langle m\rangle$, somewhat miraculously, without need to measure the actual grain positions, momenta, or contact forces.

In this paper we now ask about the nature of the microstates that causes clogging.  For $s$ spatial dimensions, what is it about the positions, momenta, and/or contact forces of the $\mathcal O(D/d)^{s}$ grains in the hole region that leads to a clog?  In principle this could be addressed by simulation, where these microscopic quantities are all perfectly known.  As a different approach, we measure and compare clogging behavior versus hole size for experimental systems that are identical but for one major difference: In one the grains are in air and have collisional/inertial dynamics, and in the other the grains are totally submerged in water and have overdamped viscous dynamics as well as reduced friction.  Once a clog forms, the stability criteria for the grains in the arch/dome are the same; however, the dynamics of arch formation must be very different.  As shown below, we find that the clogging statistics are not strongly affected.  Therefore, we conclude that grain positions are key to predicting clogging probabilities.


\section{Materials and methods}

\begin{figure}[th]
\includegraphics[width=3in]{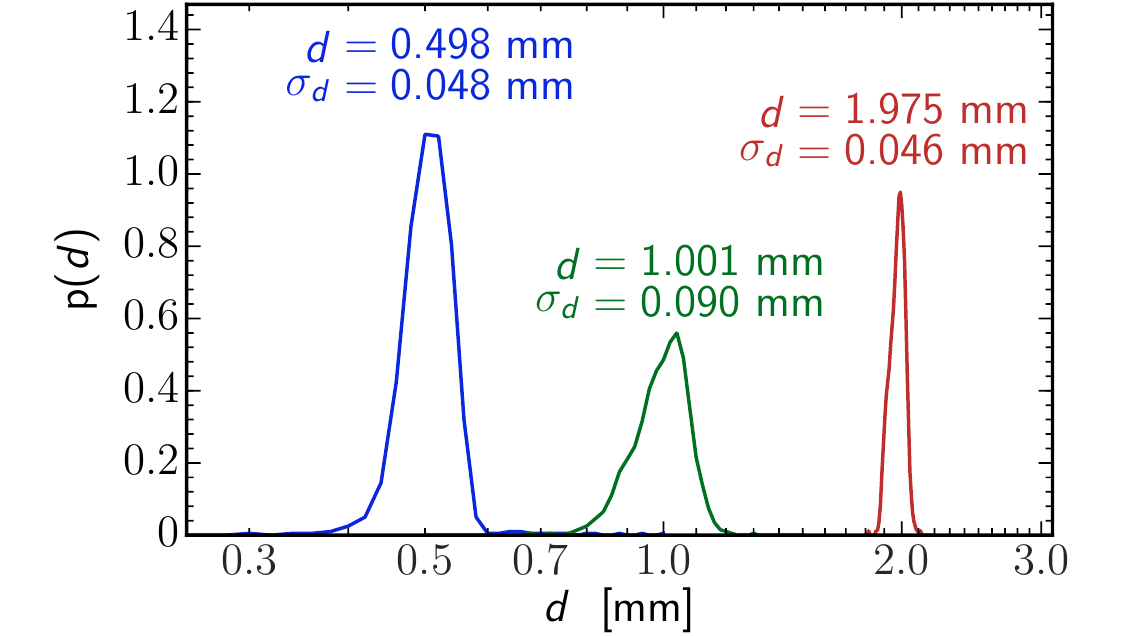}
\caption{Normalized distribution of particle diameters for glass beads, measured by a Retch Technology Camsizer and labeled by average and standard deviation.}
\label{size_distributions}
\end{figure}

The experimental granular system consist of three sizes of technical quality glass beads (Potters Industries A-series) with material density $\rho_g=2.54\pm0.01~\mathrm{g/cm^3}$.  The grain diameter distributions are measured using a Retch Technology Camsizer.  Results are displayed in Fig.~\ref{size_distributions} along with the mean, $d$, and standard deviation, $\sigma_d$.  As shown, the grains have a $5-10\%$ polydispersity, and will be referred to by their nominal diameter values of $d =$~0.5, 1.0 and 2.0~mm.   Twenty to thirty percent of the $d=1.0$~mm beads have multiple sharp edges, by visual inspection. The other beads can be described as round. This does not seem to affect the clogging results found below, and neither does the larger relative polydispersity for the $d=1.0$~mm grains. Nevertheless, the clogging analysis is highly sensitive to the size of the particles relative to the hole, in light of Eqs.~(\ref{mexp}-\ref{mdiv}).  In both dry and submerged cases the grain volume fraction is measured to be $\phi=0.58\pm0.04$, which is close to expectation \cite{FarrellSM10}.  Therefore, the mass density of the packing is $\rho=\phi\rho_g=1.47\pm0.10~\mathrm{g/cm^3}$.  The draining angle of repose is about $24^\circ$ when the grains are dry, and about $21^\circ$ when they are fully immersed in water \cite{Wilson2014}.

In the dry experiments, air conditions are controlled by standard laboratory air handling with humidity ranging between $20 - 50$~rH and temperature between $20 - 25$~C. In submerged experiments, the fluid is filtered tap water with standard textbook properties: density $\rho_f = 1.00\pm0.01~\mathrm{g/cm^3}$ and viscosity $\eta=1.00\pm0.01~\mathrm{mPa}$.  With these system parameters, the Reynolds number based on bead size and single-grain terminal falling speed in water are ${\rm Re} = \rho_f v_t d/\eta = \{40, 150, 500\}$ for the three grain sizes.  The Stokes number is ${\rm St} = (1/9) \rho_g v_t d/\eta$, which here is about Re/3.  It refers specifically to grain inertia \cite{FarrellSM10}, the importance of which is overestimated by using $v_t$ since the discharge speed is smaller in water than in air \cite{Wilson2014, SandsOfTime} and also since the relative speed of neighboring grains in the coarse-grained flow field is smaller still.  Even so, Ref.~\cite{FarrellSM10} shows that sedimenting grains need to have ${\rm St}>30$ in order for their inertia to jar a loose packing into a dense packing.  Therefore, we can expect very different dynamics for submerged versus dry grains.

\begin{figure}[ht]
\includegraphics[angle=90,width=2.5in]{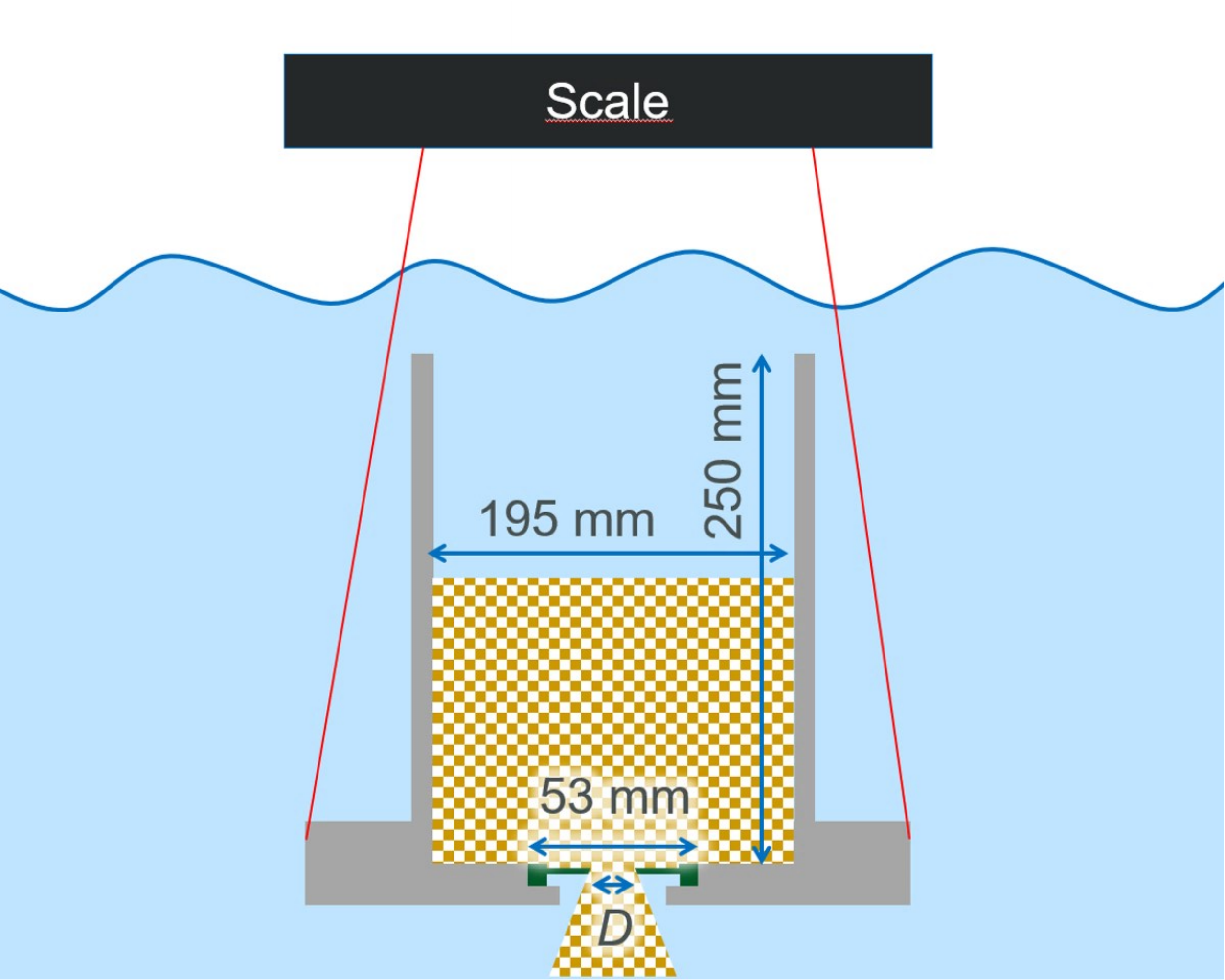}
\caption{Schematic illustration of the clogging apparatus, shown in vertical cross section. The hopper (gray) is cylindrical and hangs from a digital balance (black). An orifice with adjustable diameter, $D$, fits into a depression in the bottom plate. The walls and bottom plate are made of polycarbonate, respectively 6~mm and 13~mm thick. The flowing grains are indicated with brownish shading. The dry case is identical, except that the fluid (light blue) is absent.  Note that the top of the hopper is open, so that there is no back-flow of air or water into the hopper as the grains exit.}
\label{schematic}
\end{figure}

The apparatus for clogging measurements is shown schematically in Figure \ref{schematic}, and is similar to that in Ref.~\cite{SandsOfTime}.  The hopper consists of a flat bottomed cylinder with inner diameter $D_h = 195\pm0.5~\mathrm{mm}$ and height $h = 250\pm1~\mathrm{mm}$.  The top is open, and is completely underwater for the submerged experiments. The bottom has a depression that is fitted with an adjustable iris that serves as the hole through which the grains exit.  The hole diameter is measured with a caliper, and is circular to within $\Delta D = 0.1~\mathrm{mm}$. This is the largest source of uncertainty, especially for small orifice diameters. Therefore we also perform additional runs where the iris is replaced by an Aluminum disk with a precision-machined hole, to rule out systematic errors.  The entire hopper hangs from a digital balance (Ohaus Valor 7000) that records the change in mass with 10~Hz frequency and 1.0~g repeatability.  Alternatively, for avalanches smaller than about 10~g, the grains are collected in a cup that is weighed with a more accurate balance (Ohaus Navigator, 0.1~g repeatability). 

When a clog forms, the flow is re-started using one of the three methods. In the submerged case, a stream of water is directed underneath the hole in order to break the clog.  Alternatively, the clog is broken by poking it with a stick either manually or via stepper motor. This is used in all the dry cases, and in some of the submerged cases.  For these two methods, the water pump and the stepper motor are connected to the same computer that is interfaced to the balance, so that the system is fully automated as in Ref.~\cite{MazaPRE05}.  A third method is to manually tap to the side of the hopper.  This is useful for the $d=0.5~\mathrm{mm}$ grains, where the orifice size can be less than 2~mm in diameter and hence difficult to poke.  In all cases, the procedure is to initiate flow, to measure the total mass of grains that is discharged before a clog forms, and to repeat as desired for a large number of discharge events.  No difference in behavior was noticed for the three unclogging methods or the two hole types.


\section{Discharge Mass Statistics}

The first question is whether or not clogging remains a Poisson process when the grains are submerged.  In particular, the distribution of discharge masses is exponential in air and vacuum, but could very well be different under water.  To investigate this, as well as the fraction $F$ of flow microstates that cause a clog, we measure the discharge masses for a large number of flow events, for all three grain sizes, and for hole sizes ranging from slightly larger than one grain to as large as was reasonably feasible.  The upper limit on hole size is such that the average discharge mass is $\mathcal O(1000~{\rm g})$, which is ten times smaller than the capacity of the hopper.  This allows sampling of long-duration events without need for re-filling, which is infeasible for the submerged cases.   In dry cases, the upper limit is also set by the accuracy with which we could change the hole size.  Only a very slight increase in hole diameter is needed to increase the average flow duration from several hours to several days or weeks, i.e.\ from barely feasible to not possible.  Altogether for dry and submerged cases, and for the three grain sizes, we examined 46 different hole sizes and thousands of discharge events. The goal of the experimental procedure was to measure at least 30 discharge events for each combination of $D$ and $d$. The average number of events is 104 for each measurement point. Half of the measurements have more than 52 events and the largest measurement consists of 1172 events.  There are only three measurement points with less than 10 discharge events, all for very long duration (nearly  infeasible) runs at the largest $D/d$ ratio.

\begin{figure}[ht]
\includegraphics[width=3in]{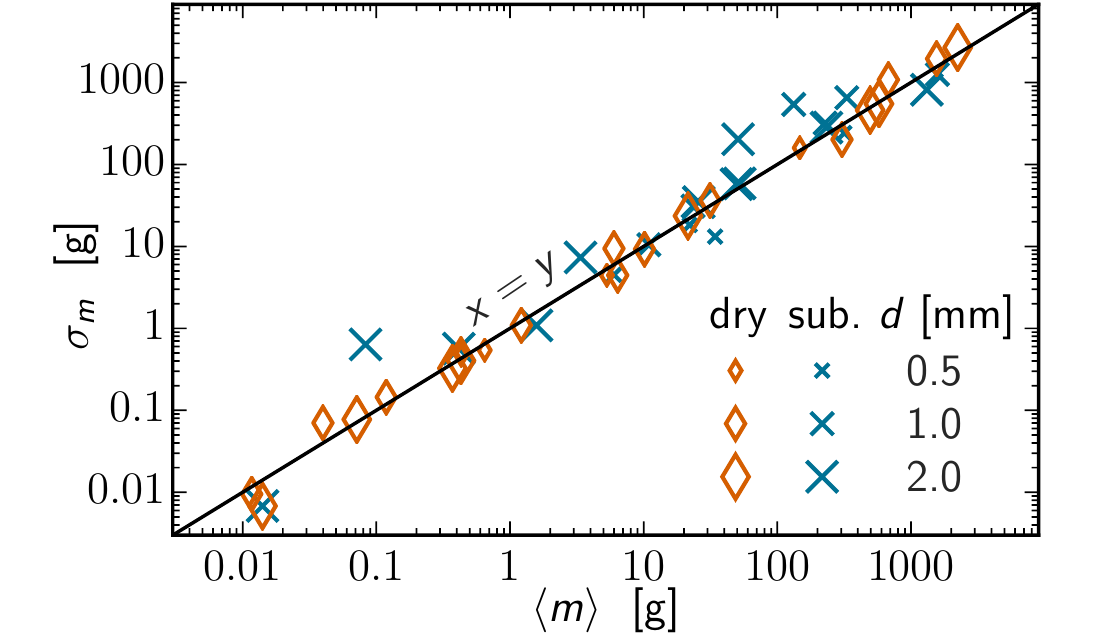}
\caption{Standard deviation versus mean for the distribution of discharge masses, for all measured combinations of hole and grain sizes, under both dry and submerged conditions.  The data fall on the line $y=x$, which implies that the distributions are exponential and that clogging is a Poisson process.  This figure has two less data points than seen in later figures, where $\langle m\rangle$ but not $\sigma_m$ were measured by collecting multiple events into a cup and weighing.}
\label{mean_std}
\end{figure}

\begin{figure}[ht]
\includegraphics[width=3in]{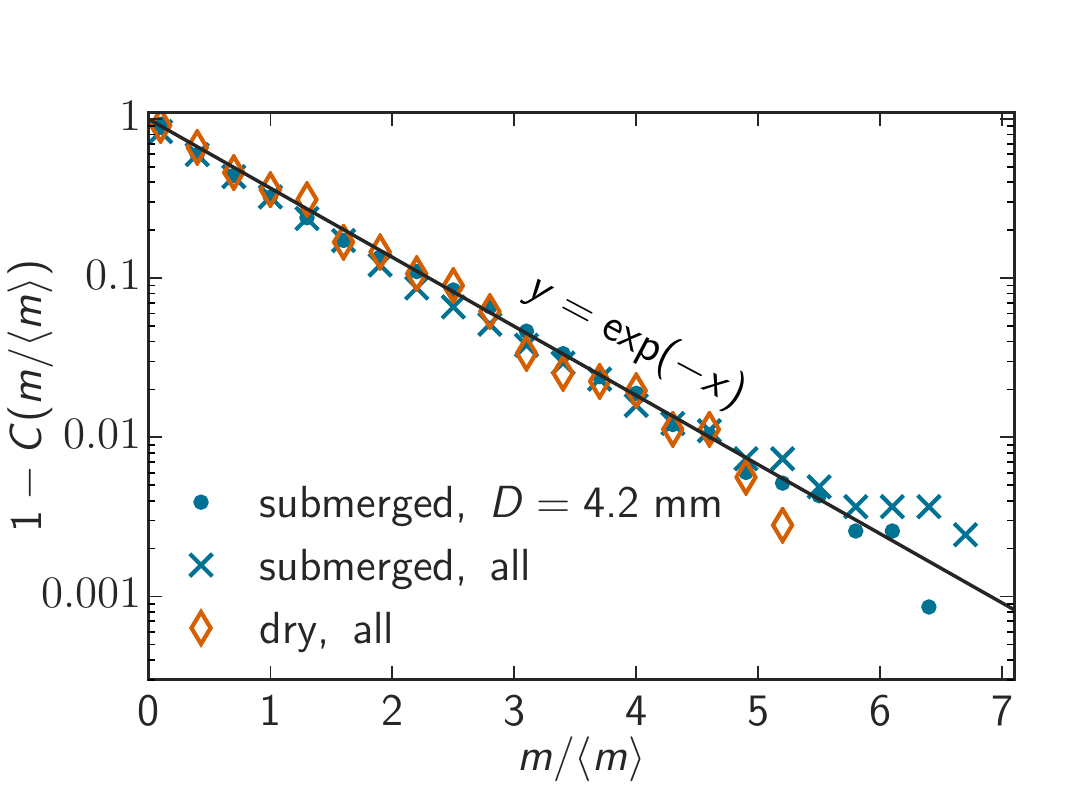}
\caption{One minus the cumulative distribution function versus scaled discharge event mass for $d=1$~mm diameter grains.  Solid circles represent the submerged dataset with the largest number of events (1172); crosses and diamonds represent all datasets for submerged and dry cases, respectively.   The solid line labeled $y = \exp(-x)$ is the expectation for an exponential distribution of discharge masses; its good agreement with the data demonstrates that clogging is a Poisson process.}
\label{cumulative}
\end{figure}

To reveal the nature of all the discharge distributions, we compute both the average $\langle m\rangle$ and the standard deviation $\sigma_m$ of the discharged masses, for a given set of conditions, and plot one versus the other in Fig.~\ref{mean_std}.  The data points fall on the line $\sigma_m=\langle m\rangle$, where $\langle m\rangle$ varies by almost six orders of magnitude as the hole and grain sizes are changed.  This is consistent with an exponential distribution.  In addition, we also collected greater statistics for a couple of specific grain / hole size combinations and directly confirmed that the distributions are nearly exponential.  In Fig.!\ref{cumulative}, we plot the cumulative distribution for the one representative set with a large number discharges and the combined submerged and dry cases for $d=1$ mm grains. The combined sets are first normalized with average $m$ before combining. The distributions are linear in semilog scale over a wide range, indicating that indeed we have exponential behavior. Thus, we conclude that clogging is also a Poisson process for submerged grains.  This is the first such demonstration, to our knowledge.  Importantly, it permits us to analyze $\langle m\rangle$ in terms of $F$, below.

\begin{figure}[ht]
\includegraphics[width=3in]{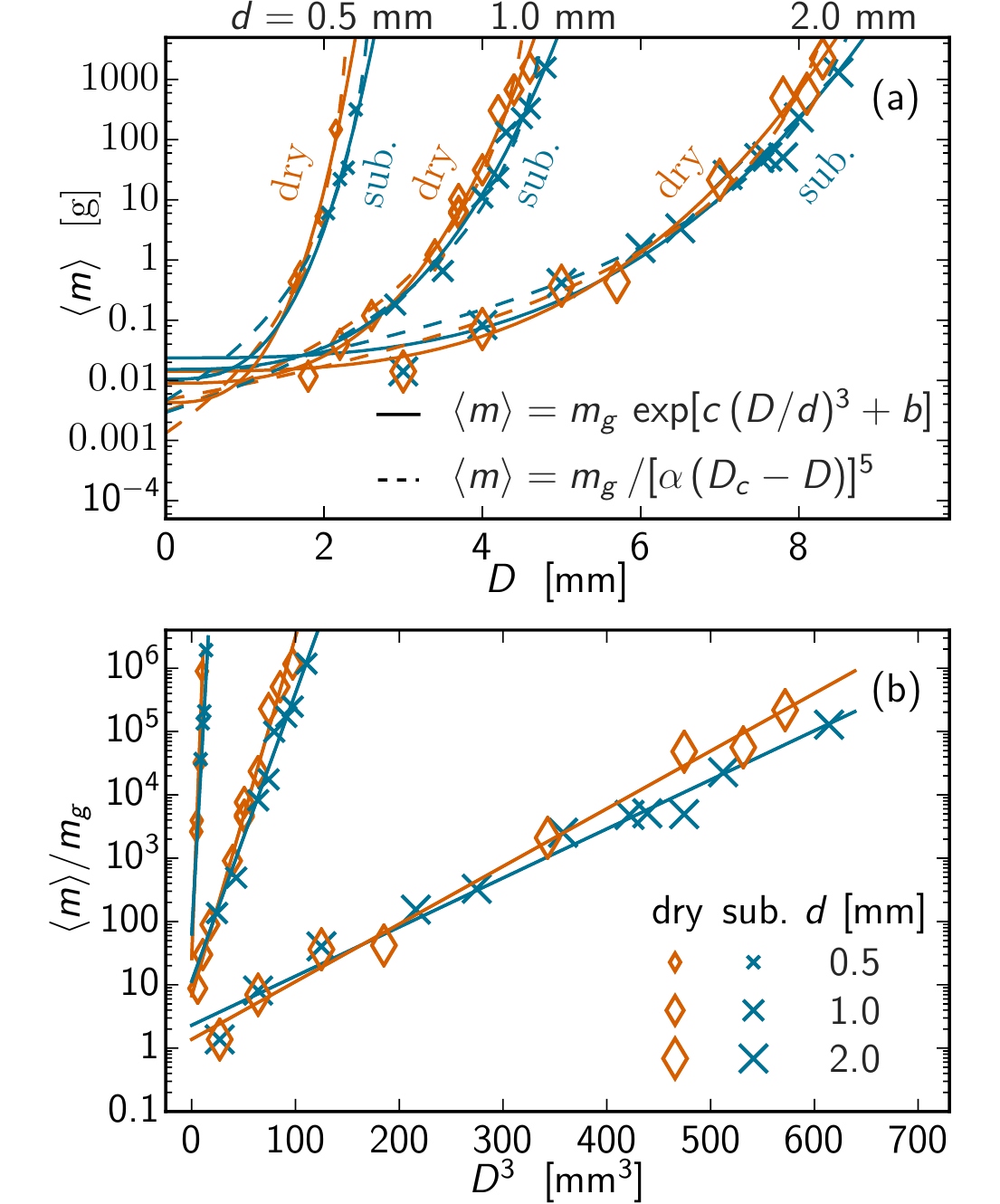}
\caption{The average discharge mass versus hole diameter (a), and scaled by grain mass and plotted versus the cube of hole diameter (b).  In both plots, the solid curves represent fits to Eq.~(\ref{mexp}) while the dashed curves represent fits to a diverging power law Eq.~(\ref{mdiv}) with exponent taken to be $\gamma=5$.  The former is exponential in $D^3$ and hence comes out as a straight line in the bottom plot.  Fitting parameters are given in Table~\ref{tab:meanParameters}.}
\label{exp_power}
\end{figure}

The next question concerns how the average mass varies with hole diameter.  To investigate, results for $\langle m\rangle$ versus $D$ are plotted two different ways in Fig.~\ref{exp_power}.  The top plot is a log-linear version of the raw data.  For the dry grains, it shows a very rapid increase that can be well-fit by both the exponential and diverging power-law forms, Eqs.~(\ref{mexp}-\ref{mdiv}), as expected.  Fitting parameters are collected in Table~\ref{tab:meanParameters}.  Also as expected, doubling the grain size requires doubling the hole size to achieve the same average discharge mass.  For submerged grains, our new result is that the average discharge mass is slightly reduced, compared to the dry case at a given hole size.  Furthermore, the functional form appears to be unaltered, in that good fits are also obtained using Eqs.~(\ref{mexp}-\ref{mdiv}).  To highlight the exponential form, the average mass data are scaled by the grain mass $m_g$ and are shown as a log-linear plot versus $D^3$ in Fig.~\ref{exp_power}b.  This causes the data to fall onto straight lines, which is the expectation for the Eq.~(\ref{mexp}) form that grows exponentially in $D^3$.  Note, too, that these fits extrapolate close to $\langle m\rangle/m_g=1$ as the hole size decreases toward zero: For the smallest holes, only a few grains escape before a clog forms.  

As further remarks on fitting, first note that the diverging power-law fits in Fig.~\ref{exp_power} assume the exponent to be $\gamma=5$.  This value is taken from  Ref.~\cite{Thomas2013}, and is roughly in the middle of the range of values reported by others.  Similarly good fits can be obtained by adjusting $\gamma$ at a fixed critical hole size of $D_c=5d$ for dry grains and 10\% larger for submerged grains.  Either way, the uncertainties in fitting parameters are quite large (and larger than for $\{c, b\}$ in the exponential fits).  Even better-looking power-law fits can be obtained by adjusting all three parameters, \{$D_c$, $\gamma$, $\alpha$\}; however, the parameter uncertainties are unacceptably large.  Similarly, the exponent may be adjusted in the form $\langle m\rangle \propto \exp[c(D/d)^s]$; however, the fitted values are close to 3, which is expected based on the model of Ref.~\cite{Thomas2015} where $s$ naturally equals the number of spatial dimensions.  Overall, the relative quality of the two fitting forms is comparable, but the smaller uncertainties and the clear physical meaning of the exponent point in favor of the exponential form.  In consequence, we reinforce the belief that there is no sharp clogging transition, i.e.\ that all granular hoppers are susceptible to clogging (though perhaps with unobservable probability).

\begin{table}
\caption{Parameters for the fits displayed in Fig.~\ref{exp_power} to Eqs.~(\ref{mexp}-\ref{mdiv}), obtained using the Fortran ODRPACK algorithm \cite{Boggs1989}.  Error estimates are given by the square root of diagonal elements of the covariance matrix. For Eq.~(\ref{mexp}), the exponent is fixed to $\gamma=5$.}
\begin{ruledtabular}
\begin{tabular}{cccc}
	 & $d$ (mm)& dry & submerged \\
	 \hline\rule{0pt}{2.5ex} 
	 & 0.5 & $0.13\pm0.03$ & $0.09\pm0.03$ \\
   $c$ & 1.0 & $0.13\pm0.01$ & $0.11\pm0.03$ \\
	 & 2.0 & $0.16\pm0.01$ & $0.14\pm0.01$ \\
	 \hline\rule{0pt}{2.5ex} 
	 & 0.5 & $10\pm2$ & $11\pm3$ \\
   $b$ & 1.0 & $8.8\pm0.6$ & $9\pm2$ \\
	 & 2.0 & $7.2\pm0.6$ & $7.7\pm0.6$ \\
         \hline\rule{0pt}{2.5ex}
	       & 0.5 & $2\pm1$ & $3\pm2$ \\
    $D_c$ & 1.0 & $4.8\pm0.8$ & $5.0\pm0.9$ \\
	       & 2.0 & $9\pm1$ & $9\pm0.6$ \\
	       \hline\rule{0pt}{2.5ex} 
	       & 0.5 & $0.03\pm0.09$ & $0.02\pm0.10$ \\
$\alpha$ & 1.0 & $0.05\pm0.04$ & $0.04\pm0.05$ \\
	       & 2.0 & $0.07\pm0.05$ & $0.06\pm0.04$
\label{tab:meanParameters}
\end{tabular}
\end{ruledtabular}
\end{table}


\section{Analysis of Flow Microstates}

\begin{figure}[th]
\includegraphics[width=3in]{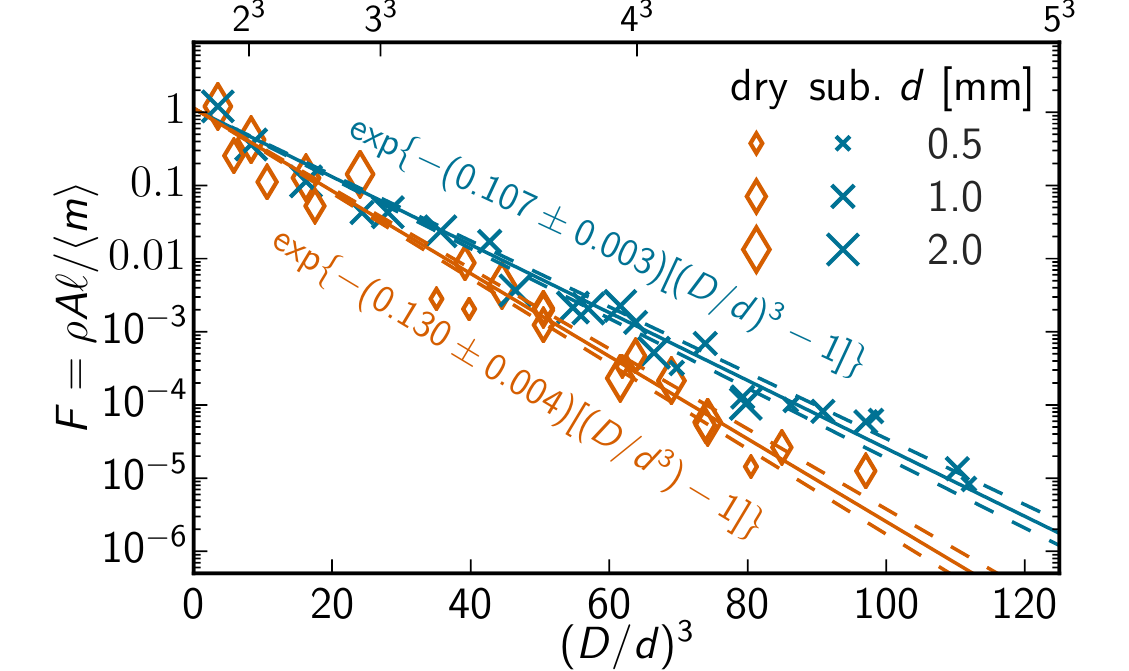}
\caption{The fraction of flow configurations that cause a clog versus the cube of hole diameter divided by grain diameter.  Experimental results are for three grain sizes, and under both dry and submerged conditions.  The solid lines represent fits to $F=\exp\{-C[(D/d)^3-1]\}$, and the dashed lines represent the range of fitting functions given by the quoted value and uncertainty in the fitting parameter $C$.}
\label{fraction}
\end{figure}

We now use Eq.~(\ref{FfromM}) to analyze the average discharge mass data in terms of the fraction $F=\rho A\ell/\langle m\rangle$ of flow microstates that precede, i.e.\  that {\it cause}, the formation of a stable clog.  In this expression, all quantities on the right-hand side are known from the measurements discussed above except for the sampling length, $\ell$.  This is the average downward displacement of grains in the hole region that is needed to create a new configuration and hence a new chance to clog.  We take it to be $\ell=(0.75\pm0.20)d$, as measured by two methods in Ref.~\cite{Thomas2015}.  The resulting behavior for $F$ versus $(D/d)^3$ is shown by log-linear plot in Fig.~\ref{fraction}.  Note that this causes the data to collapse onto two straight lines, one for dry grains and one for submerged.  These decay rapidly, since the susceptibility to clogging decreases dramatically with increasing hole size.  Both cases may be fit to $F = \exp\{-C[(D/d)^3-1]\}$, which has the correct form and is also correctly normalized to $F=1$ at $D=d$.  By adjusting only the decay rate constant, $C$, we obtain very good fits as shown.  The fitting uncertainty is about 3\%, and the decay constant for the dry grains is about 20\% larger than for the submerged grains.

Since the decay of $F$ is faster for the dry grains, they are slightly less susceptible to clogging.  This is counter to our initial expectation, which was that lubrication forces between approaching grains and reduced friction between contacting grains (vis-\`a-vis the smaller repose angle) would both render submerged grains less susceptible to clogging.  This points to grain inertia, which has a destabilizing effect on arch formation and is much larger for the dry grains.  Intuitively, to form a stable clog, an incipient arch much be strong enough to withstand collision from the grains colliding with it from above.  Of all possible arches, fewer can be stably formed in air because they must be stronger.  Conversely, a greater variety of arches can be stably formed under water since weaker ones are additionally allowed, rendering submerged grains more susceptible to clogging.  This ties in with the conclusion of Ref.~\cite{FarrellSM10} that the Stokes number controls the volume fraction of random loose packings, such that looser more delicate packings may be formed when grain inertia is absent.  This also ties in with the intuition of Ref.~\cite{ZuriguelSM2016} that incipient arches must be strong enough to dissipate the kinetic energy of the grains raining down from above.


\section{Conclusions}

In summary, we have systematically measured clogging statistics for grains being discharged from submerged hoppers, and compared them with identical but dry experiments.  We find that immersing the grains does not affect the Poissonian character of clogging, and it leads to a slightly enhanced susceptibility of clogging.  Our data reinforce the notion that a sharp clogging transition does not exist, i.e.\ that all hoppers may eventually clog given sufficient time.  Our analysis demonstrates the utility of interpreting the average discharge mass in terms of the fraction $F$ of flow configurations that cause clog formation \cite{Thomas2015}.  In particular we find that $F$ decays exponentially in $(D/d)^3$, which is roughly the number of grains in the hole region that must cooperate in order to form a stable arch (dome, really) across the hole, for both dry as well as submerged grains.  The decay rate is about 20\% slower for the submerged grains, reflecting the increase in the number of flow configurations that can form a stable clog.  Since this change is not great, we conclude that grain positions play a far more important role than grain momenta.  Due to the sign of the effect, we also conclude that it cannot be due to lubrication or friction forces.  Rather, grain inertia has some limited capacity to break incipient arches in the dry case, and this is totally removed for the submerged grains making them slightly more susceptible to clogging.  Though this picture is physically intuitive and consistent with Refs.~\cite{FarrellSM10, ZuriguelSM2016}, it is still somewhat speculative since it assumes that the position microstates during flow are unaffected by immersion in water.  This could be tested by computer simulation, or perhaps by experiments in a quasi-two dimensional geometry.

\begin{acknowledgments}
This work was supported by the Finnish Foundation's Post Doc Pool, Wihuri Foundation and Finnish Cultural Foundation (JK) and by the NSF through Grant No. DMR-1305199 (DJD).
\end{acknowledgments}

\bibliography{ReferencesHopper}

\end{document}